\documentstyle[prl,aps,twocolumn]{revtex}
\begin{document}
\draft
\tighten
\title{Photoassisted sequential resonant tunneling through
 superlattices.}
\author{Jes\'us I\~{n}arrea and  Gloria Platero \\}
\address{Instituto de Ciencia de Materiales (CSIC) and  Departamento de
Fisica de la Materia Condensada C-III, Universidad Autonoma,
Cantoblanco, 28049 Madrid, Spain.}
\maketitle
\abstract{
We have analyzed theoretically
the photoassisted tunneling current through a superlattice in the
presence of an AC potential. For that
purpose we have developed
a new model to calculate the sequential resonant current through
a superlattice based in the Transfer Hamiltonian method.
The tunneling current presents new features due to new effective tunneling
channels coming from the photoside bands induced by the AC field. Our
theoretical results are in good agreement with the available experimental
evidence.}
\pacs{73.40.G}
\narrowtext
%twocolumn
          The analysis of resonant tunneling through semiconductor
superlattices (SL) has a lot of interest from a fundamental point of view
as well as for its applications as microelectronic devices.
Their transport properties were first investigated in 1970 by Tsu
and Esaki\cite{1}.
Later, (1971), Kazarinov and Suris\cite{2} studied theoretically
the current-voltage curve
for a SL and predicted the existence of peaks corresponding to sequential
resonant tunneling between adjacent wells depending on the coupling
between the quantum wells. In 1973 Tsu
and Esaki\cite{3} presented a
calculation on resonant tunneling through multiple barriers, and
in 1974 Esaki and Chang \cite{4}observed oscillatory conductance
in a SL heterostructure.
Since then, the transport properties of those structures have been
under intense experimental\cite{5} and theoretical\cite{6}
investigation, not only under static but
also at high frequency fields.\\
When a static electric field is applied to a SL
the overlap of wave functions between wells decreases and
as a result the miniband tends to split into a series of states
which are distributed sequentially along the field direction.
In other words, a ladder of states, or
Stark ladder is formed. For high static fields
at which the voltage drop over one period is large
compared with the miniband width, coupling is completely
destroyed and the SL effectively becomes a series
of isolated quantum wells. Under these conditions,
the current-voltage characteristic has the form
of a set of narrow peaks corresponding to the
tunneling of an electron from the ground state in one cell
to an excited state in a neighboring cell (see fig. 1.a).
In reality each peak is broadened because tunneling
is even possible out of resonance due to scattering
with phonons, surface roughness etc. As a result the
current-voltage curve has a finite current between peaks
rather than zero current \cite{7,8}.\\
\par
In this paper we analyze the photoassisted sequential
resonant tunneling current through a SL in the presence
of an AC potential. The AC field
induces absorption and emission processes
which are reflected
as new features in the current density.
These new features in the tunneling
current are due to new effective
tunneling channels
coming from the photoside bands (see fig. 1.b),
which are opened and which produce additional
current steps in the characteristic current-voltage
curve. \\
In order to study the sequential resonant tunneling through a SL,
before applying the AC field, we have developed a model in
the framework of the Transfer Hamiltonian
formalism \cite{9} to calculate
 the current through the SL, considering a uniform static electric
field along the structure. In this model we have calculated first  the
probability for the electrons to cross
each individual barrier from one well
to the adjacent one,
and from that the density current. The total current
is obtained when all the individual currents result
to be equal. Under this condiction we obtain also
that all the Fermi levels in the different wells are
equal
(we are considering uniform static electric field along the SL
and neglected the electron-electron interaction effects).
The expression we obtain for the total current can be written as:
\begin{eqnarray}
J_{T}=\frac{e\hbar}{2\pi m^{*}}\frac{k_{1}(E_{11})k_{2}(E_{11})
T_{s}(E_{11})}{\alpha_{12}(E_{11})\alpha_{23}(E_{11})} \times\nonumber\\
\left[L(E_{22}-E_{11})+L(E_{23}-E_{11}) \right] E_{1}
\end{eqnarray}
where $\alpha_{ij}=(w_{b}+1/\alpha_{i}+1/\alpha_{j})$, being
$\alpha_{i}$ ($\alpha_{j}$) the perpendicular electronic wave vector in the
barriers $i$ and $j$ and  $w_{b}$ is the barrier width.
  $T_{s}(E_{11})$ is the single barrier transmission
coefficient,
$k_{1}(E_{11})$ and $k_{2}(E_{11})$ are the perpendicular component
of the electronic wave vector in the first and second well respectively,
all of then evaluated at the energy of the ground (first)
state in the first well ($(E_{11})$). $E_{ij}$ represents
the energy corresponding to the $i$ well and $j$
state referred to the conduction band bottom in the emitter.
$L(E_{2(2,3)}-E_{11})=\frac{\gamma}
{(E_{2(2,3)}-E_{11})^{2}+\gamma^{2}} $  where $\gamma$ is the half
width of the well state. Finally $E_{1}$ represents the
Fermi level for the wells and corresponds to a bidimensional
electron density. \\
\par
The carrier charge density in the wells in a SL
can be obtained by two ways: from the emitter which is
considered to be heavily doped or from
photoinjection from the valence band. Our model can be
adapted in any of the two situations.
All the scattering
processes which are responsible for the tunneling out
of resonance are taken into account in our model, through a finite
broadening in the well states which is
represented by the Lorentzian function $L(E_{2(2,3)}-E_{11})$.
We have applied this model to
analyze the tunneling current through a SL formed by 50 periods of
a GaAs well with 13.3 nm width and an AlAs
barrier with 2.7 nm width.
The wells are not doped and the carriers are provided
by photoinjection from the
valence band in order to keep constant in average the number
of electrons resposible for the current at every external dc bias.
The results are shown in figure 2 where we can see the two main
peaks corresponding to the sequential
resonant tunneling from the ground state of one well
to the first and second excited states of the adjacent one.
It can be observed how these peaks are broadened and
a finite current different from zero
between peaks which is due to the different scattering
processes that take place during the tunneling from one well
to the next one as we have explained above.
The general aspect of the current-voltage curve agrees with
the physics of the sequential resonant tunneling when the dc
electric field
is dropped uniformly across the sample and is qualitatively similar
to the ones obtained in the available experiments\cite{7,8}.
The time which describes the relaxation between the states
involved in the resonant
tunneling process due to momentum
relaxing collision is about $10^{-13}s.$\cite{4,7}.
With this time we estimate a half width for the
broadening corresponding to the well states
of $\gamma=5-10 meV$. For the curve in figure 2 we have used a
value of $\gamma=10 meV$.\\
\par
Once we have developed a model to obtain the current through a SL,
we have analyzed theoretically the
effect of an external AC field on the sequential resonant
tunneling of a SL.
We have used again the Transfer Hamiltonian formalism to obtain
the probability for an electron to cross from one
well to the next one $(i\rightarrow j)$
in the presence of a time oscillating
field (both wells are oscillating):
\begin{eqnarray}
P_{ij}=\lim_{\alpha\rightarrow 0}\frac{d}{dt} \left|\frac{1}{i\hbar}
\int_{-\infty}^{t}<\Psi_{j2}|V_{L}|\Psi_{i1}>e^{\alpha t^{'}}dt^{'}
\right|^{2}
\end{eqnarray}
where $V_{L}$ represents the potential barrier.
$\Psi_{in}$ are the resulting electronic wave functions
in the time modulated quantum well, i.e, for a Hamiltonian of
the form $H=H_{0}+eFz cos{wt}$ being $F$ the intensity for the AC
electric field, $w$ its frequency, the subscript $i$ represents the well
and $n$ represents the resonant states.
Those electronic wave functions can be written following Tien and
Gordon \cite{10}:
\begin{eqnarray}
\Psi_{in}=\Psi_{0} \sum_{-\infty}^{\infty} J_{n}(\beta_{i})e^{-inwt}
\end{eqnarray}
where $\Psi_{0}$ is the unperturbed wave function (without the AC field)
, $J_{n}(\beta)$ are the Bessel function of the first kind, and $\beta_{i}=
eFz_{i}/\hbar w$, being $z$ the spatial coordinate in the SL axis.
We consider that the AC field affects the whole structure increasing
uniformly from the emitter, which is considered to be fixed, to the
collector which is affected by the field with its total amplitude.
In between, all the wells are oscillating with increasing intensity
depending on its spatial coordinate.
 From the probability we can calculate the current
through each individual
barrier inside the SL. We obtain the total current
when all the individual currents
result to be equal in the same way as we have explained above.
Finally we have an expression for the total current which is
crossing the SL in the presence of an external AC field:
\begin{eqnarray}
J_{T}&=&\frac{e\hbar}{2\pi m^{*}}\frac{k_{1}(E_{11})k_{2}(E_{11})
T_{s}(E_{11})}{\alpha_{12}(E_{11})\alpha_{23}(E_{11})}E_{1}\times\nonumber\\\
& &\sum_{m,n=-\infty}^{\infty} J_{m}^{2}(\beta_{2})J_{n}^{2}(\beta_{1})
\times [ L(E_{22}-E_{11}+(m-n)\hbar w)+ \nonumber\\
& &(E_{23}-E_{11}+(m-n)\hbar w)]
\end{eqnarray}
where $\beta_{2,1}=\frac{eFz_{2,1}}{\hbar w}$ being $z_{2,1}$ the
spatial coordinates corresponding to the center of the second and
first wells respectively.\\
We have applied the model
described above to the available experimental evidence \cite{11}.
We have calculated the current through a SL consisting of
100 periods of 33 nm $GaAs$ wide quantum wells separated by 4 nm
$Al_{0.3}Ga_{0.7}As$ barriers. As in the paper \cite{11}, we consider the
radiation is coupled into the sample by using an antenna and produce
an electric field intensity of several $kV/cm$ . In fact we have use an
electric field intensity of $10^{5} V/m$
and a frequency of $2.11\times 10^{12} s^{-1}$
which is equivalent to a
photon energy of $\hbar w=1.39 meV$.
As in the case without AC
field, we have considered a bidimensional density for the electrons
in the wells of the order of $E_{1}=10^{11} cm^{-2}$ and a half width for the
resonant state of the order of $5meV$.
In figure 3 we present the results obtained  for the current density
versus the dc voltage for
the parameters described above. Continuous line corresponds to no
AC field present and dotted line to the AC field present.
Figure 3a presents the results involving
the first resonant condition or in other words,
the first peak corresponding to the tunneling
from the ground state of one well
to the first excited state of the next one.
In figure 3b we can see an extended view which includes
also the second peak corresponding to the tunneling
to the second excited state.
Under an AC field new steps appear in the current-voltage curve(dotted line)
and can be observed
how the main current peaks decrease in intensity referred to the case
of no AC field present. All these new features can be explained in
terms of photon assisted tunneling. The presence of the
AC field produce new channels for the tunneling coming from
the photoside bands induced by the time dependent field and because of
that electrons in the ground state of one well
can tunnel to the excited states of the adjacent well
with the absorption or emission of one or more photons.
Or in other words, the tunneling process now, can take place
between the ground state of one well and a photoside band in
the adjacent well (see fig. 1.b).
This is the reason for the positive structure in the bias regions
around the main current peaks and also for the smaller intensities of
those peaks since the electron density in the tunneling has to
be conserved for both cases, with or without AC field.
All these new features that
can be seen in the current-voltage curve are in good agreement
with the results obtained in the available experimental evidence
\cite{11}.
According to all the explained above, the new effective tunneling
channels due to the photoside bands are predicted to appear
at energies of $\pm n\hbar w$ from the resonant states. However
these channels are not reflected individually
in the current-voltage curves, where a continuous positive
plateau appear at both sides of
the main central peak. This is not observed either in the experimental
results\cite{11}. We think this is due to the small AC field frequency
we are using ($1.39 meV$), compared to the half width of the
well resonant state ($5 meV$). If we want to see
the different channels contributions individually in the
current-voltage curve,
we have to change the parameters, making the AC field
frequency bigger compared to the
half width of the state.
Using our model we have calculated the current density for the
same SL as before and in the presence of an AC field with the an
intensity of $F=1.5 10^{5}V/m$ and with a frequency of $\hbar w=5meV$.
The other different
parameter has been the half width for the resonant state which now is
of the order of $\gamma= 1meV$.
The results are shown in figure 4.a, where we can see very
clearly the individual contributions corresponding
to the new effective tunneling channels of
$\pm 2\hbar w$, $\pm 1\hbar w$ and for the two
main resonant peaks.
In figure 4.b we compared the results for the AC field present
(dotted line) versus the case with no time dependent
field present (single line). Is interesting to note in figure 4,
how the steps and plateaus from figure 3, become
real current peaks which agrees with the concept
of photoside bands that produce new effective
tunneling channels. For those new peaks,
the tunneling take place from the
ground state of one cell to a photoside
band of the next cell. \\
In conclusion, the effect of an AC field
on the tunneling current density through a SL
has been analyzed.
A new theoretical model ,based in the
Transfer Hamiltonian method, to calculate
the current in SL has been presented and has been generalized
to the case of a time dependent field
present.
The effect of the AC field on
the current density is a new tunneling mechanism which
still is resonant and sequential but based in photon side
bands induced but the presence of the time dependent field.
New features for the current have been obtained: positive steps
and plateaus that can become real current peaks
with the appropriate conditions. Our results are in
good agreement with the available experimental evidence.
Our model can be improved taken into account
the effect of the charge inside the well on the SL
potential profile which induces the oscillatory
dependence of the current with the applied static voltage due
to the formation of domains in the SL.
This is the aim of a future work.\\
\par
 This work has been supported by the Comision Interministerial
de Ciencia y Tecnologia of Spain under contract MAT 94-0982-c02-02 , by
the Comission of the European Communities under contract SSC-CT 90 0201
and by the Acci\'on Integrada Hispano-Alemana 84-B.\\
%\nonum

\begin{figure}
\caption{a) Schematic diagram of sequential resonant tunneling
of electrons through a superlattice. The potential energy drop
across a superlattice period is equal to the energy difference
between the ground state and the first excited state.
b) Sequential resonant tunneling in the presence of an AC field.
Tunneling can be produced through new effective channels induced
by the time dependent field.}
\end{figure}
\begin{figure}
\caption{Current density versus voltage for
sequential resonant tunneling through a superlattice of 50
periods of 2.7 nm of AlAs barrier and 13.3 nm of GaAs well
The half width of the resonant state is of $10meV$.}
\end{figure}
\begin{figure}
\caption{a) Current density versus voltage for
sequential resonant tunneling through a SL with the
parameters of [11], and a
half width of $5meV$, in the presence of an AC field with an intensity
$10^{5}V/m$ and a frequency of $1.39 meV$ (dotted line) and with no AC.
field present (continuous line).
b) Same as in a) for an extended view of the applied voltage.}
\end{figure}
\begin{figure}
\caption{a) Current density versus voltage for
sequential resonant tunneling through a SL with the
parameters of [11], and a
half width of $1meV$, in the presence of an AC field with an intensity
$1.5x10^{5}V/m$ and a frequency of $5 meV$.
b) Same as in a) for (dotted line)
AC present and (continuous line)
without AC applied to the sample.}
\end{figure}
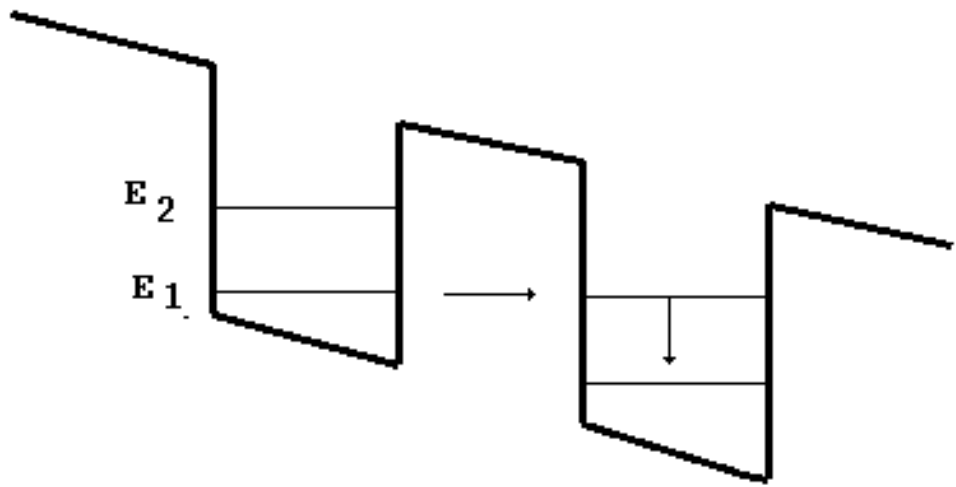
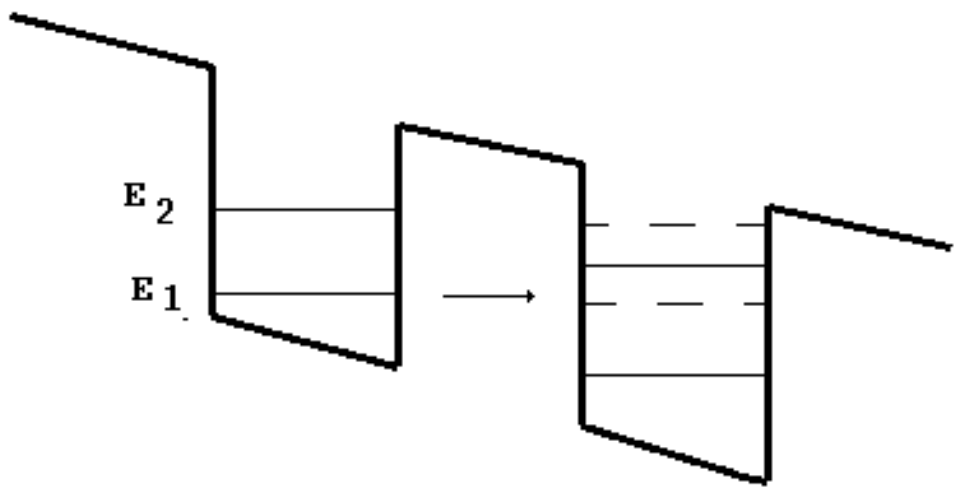
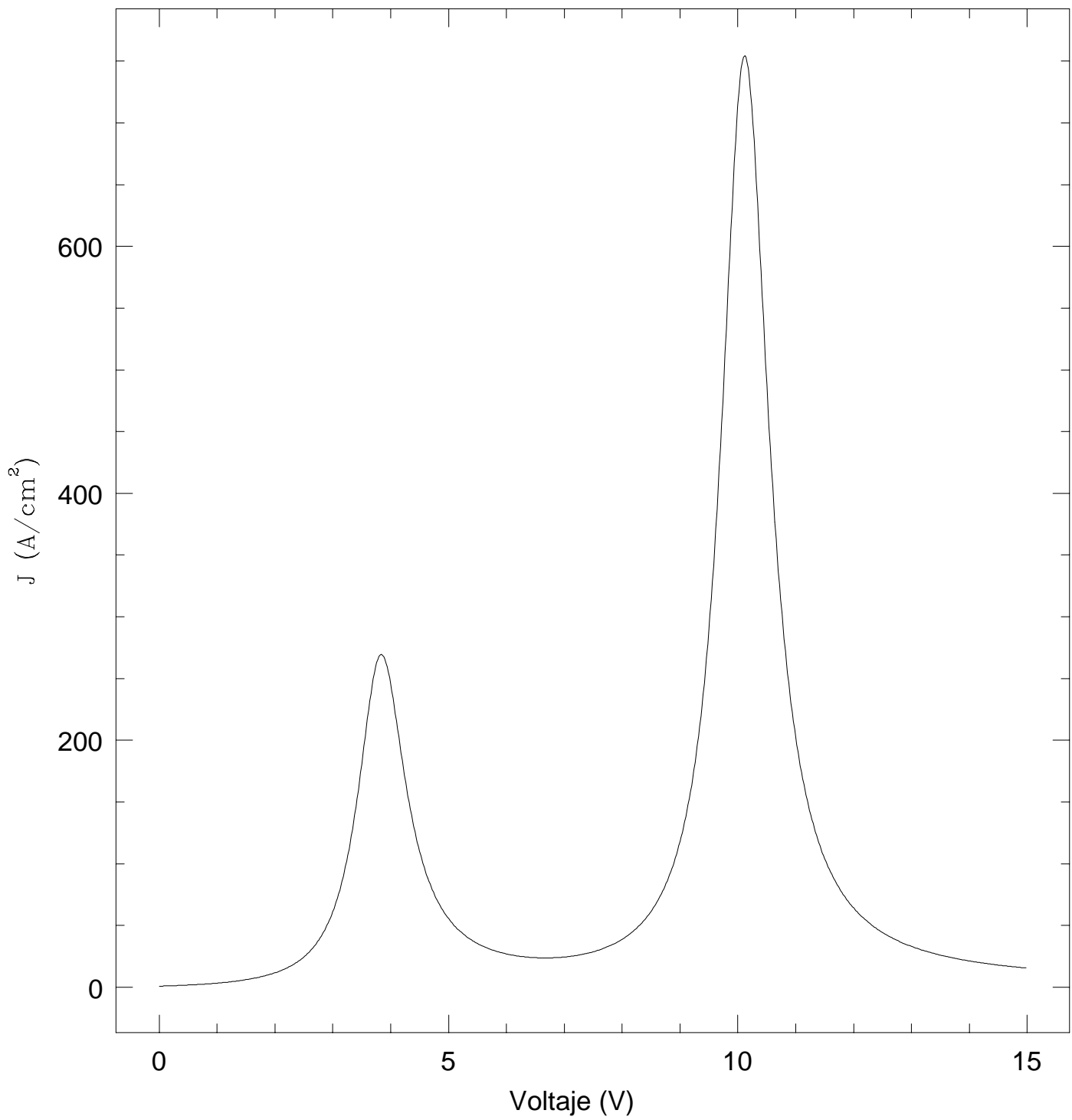
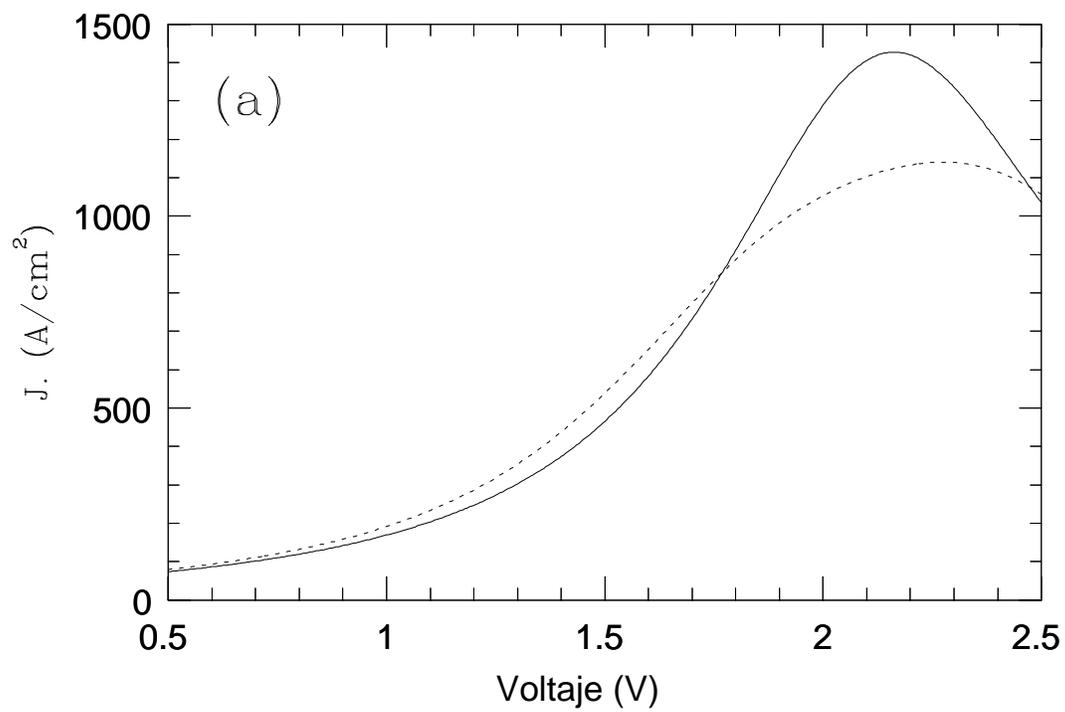
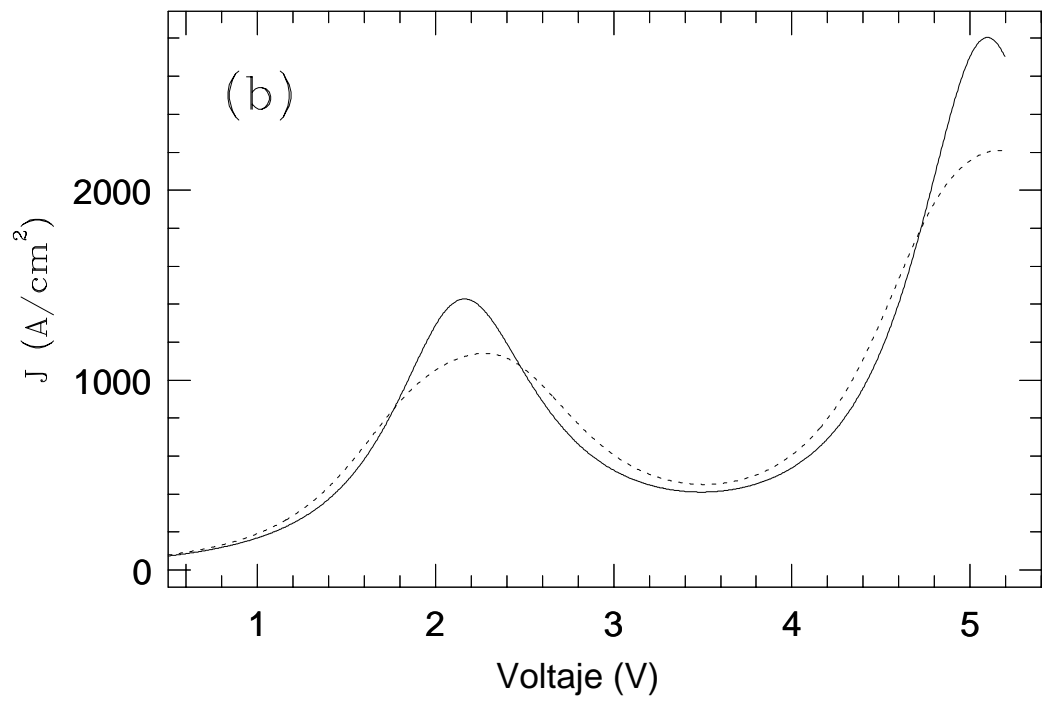
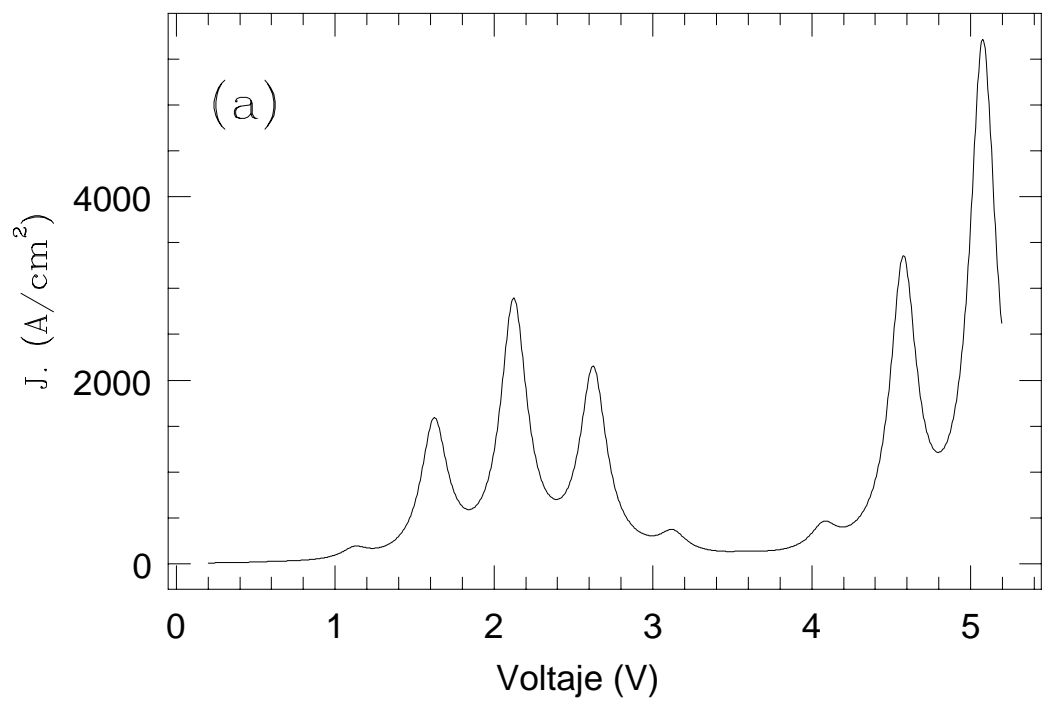
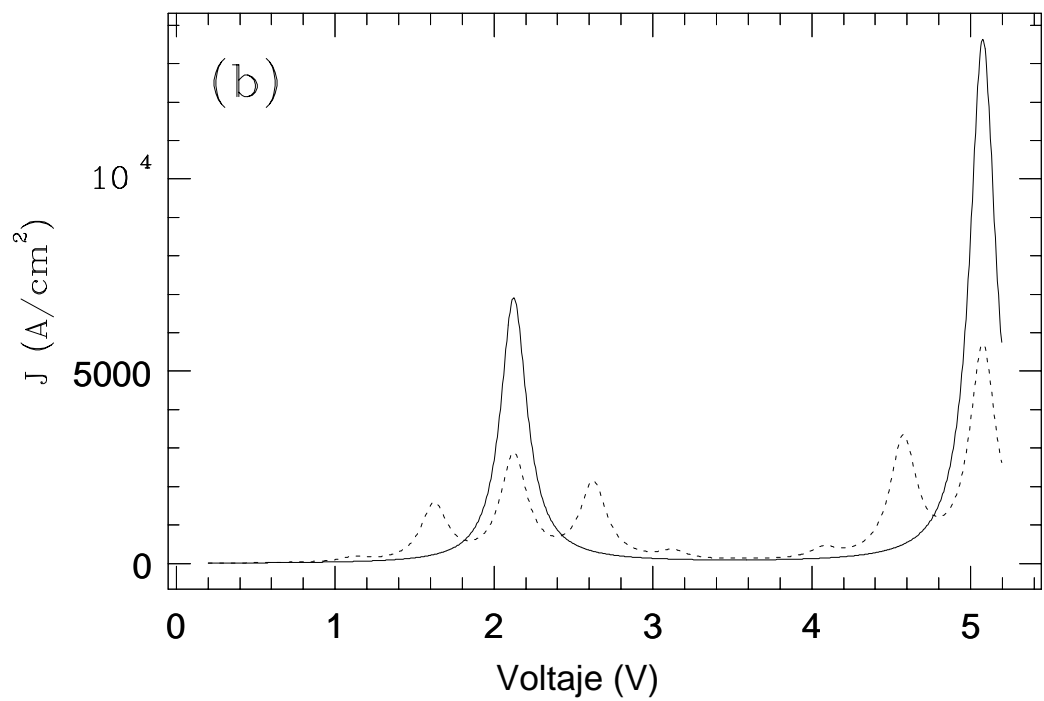
\end{document}